\documentclass[twocolumn,aps,floatfix,showpacs,superscriptaddress,amsmath,amssymb,intlimits]{revtex4}
\newcommand{\be}{\begin{equation}}
\newcommand{\ee}{\end{equation}}
\newcommand{\bea}{\begin{eqnarray}}
\newcommand{\eea}{\end{eqnarray}}

\usepackage{graphicx}
\begin{document}

\title{Anderson localization in Bose-Einstein condensates }
\author{Giovanni Modugno}
 \address{LENS and Dipartimento di Fisica, Universit\`a di
Firenze,
  and INO-CNR\\  Via Nello Carrara 1, 50019 Sesto Fiorentino, Italy}

\begin{abstract}
The understanding of disordered quantum systems is still far from being complete, despite many decades of research on a variety of physical systems. In this review we discuss how Bose-Einstein condensates of ultracold atoms in disordered potentials have opened a new window for studying fundamental phenomena related to disorder. In particular, we point our attention to recent experimental studies on Anderson localization and on the interplay of disorder and weak interactions. These realize a very promising starting point for a deeper understanding of the complex behaviour of interacting, disordered systems.

\end{abstract}

\maketitle

\section{Introduction}
Disorder is ubiquitous in nature and often, even if it is only a weak perturbation, it tends to strongly affect the properties of many physical systems. The most celebrated effect of disorder is probably Anderson localization, i.e the localization of individual particles or waves in a disordered energetic landscape. This fundamental effect of disorder has been explored in depth already more than 50 years ago \cite{Anderson58}, motivated by the obvious interest in the dynamics of electrons in crystals. There imperfections are indeed almost unavoidable, and can lead to a dramatic change of the conduction properties of real materials \cite{lee}.

Anderson localization can be understood as the effect of multiple reflections of a plane wave by random scatterers or random potential barriers. Since the amplitudes or the phases are not all identical, as they would be in a periodic potential, they tend to produce a suppression of the propagation amplitude of the wave at sufficiently large distances, even if the reflected amplitude is exceedingly small. In this case, it can be easily shown that at large distances the wavefunction must then decay as $|\phi(x)|\propto\exp{|x|/\xi}$, where the parameter $\xi$ is the localization length \cite{Mott}. Such phenomenon appears already when the potential barriers are much smaller than the kinetic energy associated to the plane wave, a region of parameters where classically one would expect the disorder to produce just a weak perturbation of the propagation.
Localization can occur not only for plane waves in free space, but also for Bloch waves, i.e. for particles or waves moving in a lattice potential. It is well known that in this case reflections at the potential barriers are already very strong even in absence of disorder, but extended Bloch states can however exist, provided that the energy of the individual lattice sites are equal or shuffled in a periodic way. In 1958, P.W. Anderson demonstrated that if the on-site energies are instead shuffled in a random way, the reflections interfere destructively and the eigenfunctions of the system become exponentially localized \cite{Anderson58}.

Much theoretical work has been done about Anderson localization in the past fifty years with important results, for example, in assessing how localization behaves in finite size systems (the so-called scaling theory of localization \cite{ahbrams79}). Several decades of experimental study of condensed-matter systems have on the one hand established the relevance of disorder to many phenomena, with recent highlights such as the quantum Hall effect \cite{qhe} or the exceptional conduction properties of graphene \cite{graphene}. On the other hand, they have not provided conclusive results about Anderson localization itself. This is mainly because in real materials there are ingredients that go beyond the approximation of a single particle in a stationary disordered potential \cite{lee,kramer93}, which make the observation of Anderson localization difficult. There one has indeed to consider also excitations of the potential (phonons) and interactions between particles, which are typically strong in electronic systems. These effects might weaken or enhance localization of particles, depending the details of the system. For example, it is well known that a repulsive elastic interaction between electrons can enhance localization \cite{kramer93}, while the same repulsion between bosons tends to screen disorder and to weaken Anderson localization \cite{lee92}. On the other hand, it is believed that inelastic processes, such as electron-phonon \cite{lee} or electron-electron \cite{altshuler06} scattering can provide a hopping energy that is sufficient to overcome localization.

To describe disordered condensed-matter systems one then needs to fully understand the interplay of disorder and interactions. This has however turned out to be an extremely difficult task not only in experiments, but also in theory \cite{kramer93}. There are currently several open questions related to this interplay, not just for electrons but even in the conceptually simpler case of bosonic systems. For example, there is still discussion in theory on how a weak interaction can affect Anderson localization \cite{Lugan07,roux08,natterman09,pikovsky08,flach09,cherroret09,paul09,wellens05}, or on the quantum phases that arise due to disorder and interactions in the regime of strong correlations \cite{Giamarchi88,Fisher88,Fisher89,Scalettar91,roux08,pilati09}.

From the experimental point of view, various engineered systems have recently appeared where transport of particles or waves can be strongly affected by controllable disorder as in the case of electrons in solids. Since most of the parameters of these systems are highly controllable, they can be employed to study in depth the basic aspects of the physics of disorder and interactions.

The first systems in which Anderson localization has been unambiguously observed (for a detailed review, see for example ref.\cite{50yearsal}) were random photonic media, where electromagnetic waves get localized by randomly distributed scatterers \cite{vanalbada85,dalichaouch91,wiersma97,storzer06}. Electromagnetic waves, as well as sound waves \cite{weaver90,maynard01,hu08}, do not suffer from the large nonlinearities of electronic systems, and the single-particle character of Anderson localization can be clearly seen. The main obstacle to the observation of localization in these systems has instead been the presence of absorption. More recently, various groups have been able to realize also photonic lattices with disorder, where various models of disorder can be tested \cite{Schwartz07,lahini08,lahini09}. These systems allow also to study the effects of interactions, thanks to presence of tunable Kerr nonlinearities. However, these nonlinearities are rather weak, and so far only the effect of attractive interactions has been clearly identified in experiments.

Localization phenomena have also been studied in the context of the quantum kicked rotor, a system that features Anderson localization in momentum space. After that seminal theoretical studies have recognized that Anderson localization has its clear signature in quantum
transport processes in periodically driven systems \cite{casati79,fishman82,casati89,schelle09} various experiments have been performed, mainly by studying the microwave ionization of atoms \cite{bayfield89,arndt91}, or ultracold atoms in free expansion interacting with pulsed periodic potentials \cite{raizen,chabe08}. The analogous of Anderson localization in 1D has been studied in depth in those systems, and recent studies have also unambiguously identified an Anderson transition in 3D \cite{chabe08}. Ongoing experiments are providing a new insight on the properties of the localization at criticality \cite{lemarie10}.

The latest systems that have appeared on the scene of disorder studies are ultracold quantum gases of atoms in traps, which are interesting in view of the unprecedented possibility of playing with both quantum statistics and of controlling the most relevant physical parameters, such as temperature, density, dimensionality and, very importantly, interactions \cite{lewenstein07,Bloch08}. In addition, using laser beams one can create potentials of almost arbitrary shape, ranging from perfectly sinusoidal lattices to random speckle potentials \cite{ovchi98}. The atomic systems can therefore be exposed in a controlled way to various kinds of disordered potentials, thus realizing several of the most interesting disordered models studied in theory. This combination of full control over disorder and interaction in a quantum system is offering new possibilities for finding a solution to some of the open questions related to the physics of disorder.

After few years of research on disorder with quantum gases, promising results have already been obtained in the context of Anderson localization (for a review, see also refs.\cite{50yearsai,Fallani08,palencia09}). Two experimental groups have reported the observation of localization of a non-interacting Bose-Einstein condensate in two different kinds of disordered potentials \cite{Billy08,Roati08}, and other studies on the interplay of disorder and interactions are on the way \cite{Chen08,deissler09}.  While these studies have not solved all the open questions about Anderson localization, they have clearly shown that a number of them can be addressed by employing atomic quantum gases. The exploration in both experiments and theory is however not limited to the weak interaction regime that is relevant for Anderson localization. For example, from the experimental point of view, notable progresses have been reported also in the understanding of the physics of strongly correlated, disordered systems \cite{Fallani07,White09}.

In this review we discuss how ultracold atomic quantum gases are being employed to investigate the physics of Anderson localization. The focus will be on the recently performed experiments on the dynamics in real space and in momentum space of Bose-Einstein condensates in correlated random potentials and in quasiperiodic lattices. We discuss both regimes of Anderson localization i.e. negligible atom-atom interactions and that in which there is an interplay of disorder and a weak interaction. We conclude by pointing out the similarities and complementarities of atomic and photonic systems in the study of Anderson localization, and by outlining the interesting research directions for the near future.

\section{Disorder models and Anderson localization}
As already outlined, at the single-particle level the term "Anderson localization" is broadly employed to name the exponential decay of the eigenfunctions of a quantum system that arises in presence of disorder. This phenomenon has of course a strong effect on the transport properties of macroscopic systems, e.g. electrons in a crystal or light in a material slab, where the Anderson localization is seen as a dramatic change of the conductance or of the diffusion constant. The details of how such localization occurs in a disordered system is of course strongly dependent on the type of disorder and on the energy spectrum that the system would have without disorder. Many models of disorder that are relevant to real systems have been studied in the past; their list is too long to be even summarized here, but there are extensive works on the subject, such as the theoretical review presented in Ref. \cite{lifshits88}).

A disordered system that is often considered in theory consists of a free particle experiencing a series of uncorrelated random potential barriers. The barrier heights are supposed to have a continuous distribution, with no lower bounds. This corresponds to have a $\delta$-like correlation function for the potential, $g(z)=\langle V(x)V(x+z)\rangle=\Delta^2\delta(z)$. Throughout this work the quantity $\Delta$ will represent the mean potential energy shift associated to disorder, or disorder strength. For this special case one finds that in an infinite 1D system all the eigenstates are always localized, even for exceedingly small values of the average barrier height. This is a consequence of the fact that in 1D each barrier causes a complete reversal of the motion of a part of the incident wave. The localization length $\xi(E)$ is however approximately linear in the kinetic energy, as one would intuitively expect since the effect of the quantum reflections decreases as $E$ is increased much above the potential barriers. In 1D it is possible to establish a series of fundamental relations that help in understanding the properties of the disordered systems. For example, a relation between the localization length and the density of states $\rho(E)$ that can be useful in many instances has been demonstrated by Thouless \cite{thouless74}
\begin{equation}
\xi(E)^{-1}=\int^{+\infty}_{-\infty}\ln|E-\epsilon|d\rho(\epsilon)\,.\label{eq:thouless}
\end{equation}
Also, in the limit of large energy compared to the disorder, $E>\Delta$, another relation between $\xi$ and the correlation function can be derived \cite{lifshits88}
\begin{equation}
\xi(E)^{-1}=\frac{1}{8E}\int^{+\infty}_{-\infty}g(z)dz=\frac{\Delta^2}{4E}\,.\label{eq:lifshits}
\end{equation}
For the disordered potential above this gives $\xi(E)\propto E/\Delta^2$. Note that in a random potential the case $E<\Delta$ must be treated separately, since there is also the possibility of having "classical" bound states in the few deepest wells, in the so-called Lifshits tail of the energy distribution \cite{lifshits88}.

An analogous disordered potential in 2D gives rise to a similar picture, i.e. the states are always localized in an infinite system, but the localization length now increases exponentially with the energy, $\xi(E)\propto \exp(E^{1/2})$. This different behaviour can be understood in a random-scattering picture as a consequence of the larger number of possible paths in 2D, or simply of the fact that a reflection process doe not necessarily imply a change of sign of the velocity.  For the 3D case, theory indicates that the states are not necessarily always all localized. For a finite disorder strength there exists indeed a phase transition between extended and localized states at a specific energy $E_m$, the so called mobility-edge. The eigenstates of the system are localized only for $E<E_{m}$, while they are all extended for $E>E_{m}$. For a given system, the position of the mobility edge depends on the disorder strength. Note that the determination of $\xi(E)$ is instead more complex in higher dimensionality than in 1D, and often one can only rely on numerical calculations \cite{kramer93,schreiber95}.

As we will outline in the following, the results found for this ideal case change when one considers particles with different dispersion relations, e.g. particles in a lattice, or more realistic correlated disordered potentials. In the rest of this review we will concentrate our attention to the 1D case that is more relevant to the experiments performed so far with Bose-Einstein condensates.

\subsection{Random disorder in free space}
A type of disorder that is close to the random potential disorder above can be created for Bose-Einstein condensates by employing laser speckles \cite{lye05,bouyer10}. The quantum gas is initially prepared by evaporative cooling in a smooth harmonic potential. Random disorder is then added separately by employing the static dipole potential of a laser beam that has passed through a diffusive plate and has been then focused onto the atomic sample. The plate creates a random phase shift along the beam profile, which is converted into an intensity distribution in the far field that has randomly distributed peaks. If the laser wavelength is for example smaller than the wavelength of the relevant atomic transition, the corresponding dipole potential is repulsive, and such intensity peaks act as potential barriers for the particles, as shown in Fig.1a.
Supposing that the disorder is present only along one of the axes of the harmonic potential, one has to deal with a 1D problem described by the Hamiltonian
\begin{equation}
H=-\hbar^2\nabla^2/2m+\frac{1}{2}m\omega^2x^2+V(x)\,. \label{eq:random}
\end{equation}
The relevant information about the disordered potential is gained by looking at the energy distribution, which is exponential, $P(E)\propto \exp(-E/\Delta)$ and at its correlation function $g(z)=\langle V(x)V(x+z)\rangle$. Supposing that the lens that creates the speckle has a a focal length $f$ and a diameter $D$, a good approximation for the correlation function is
\begin{equation}
g(z)=\Delta^2\frac{\sin^2(z/\sigma_r)}{(z/\sigma_r)^2}\,,\label{eq:sanchez1}
\end{equation}
where $\Delta$ is the mean height of the barriers and $\sigma_r=\lambda f/\pi D$ represents the correlation length of the potential. The presence of a non-zero correlation length must be considered with some attention, since it changes substantially the localization properties of the system introducing an effective mobility edge, i.e. it allows the presence of extended states even in 1D \cite{izrailev,kuhl}. As an order of magnitude, for a diffraction-limited system one has $\pi\sigma_r\approx \lambda$, where $\lambda\approx$1$\mu$m is the wavelength of the laser used to create the speckles.
How to intuitively understand the effect of a finite correlation length? As we already mentioned, the localization of a plane wave arises from the destructive interference of the reflection from individual potential barriers.  Now, we can imagine that if a plane waves has a wavelength smaller than $\sigma_r$, it will not feel the barriers like steep potential edges but more like smooth variations of the potential, therefore with a vanishingly small reflection amplitude. If this is the case, the plane waves with energy $E=\hbar^2k^2/2m$ such that their wavevector is $k>1/\sigma_r$ cannot be strongly localized by disorder. This point can be seen in a more formal way by employing the Wiener-Kintchine theorem, that establishes a relation between the correlation function and the momentum components of the disordered potential \cite{izrailev}. Theoretical studies \cite{sanchez07,gurevich09} have indeed found that for a realistic speckle potential there's an effective mobility edge at an energy $E_m=\hbar^2/2m\sigma_r^2$, and the appropriate form of the localization length within the Born approximation is
 \begin{equation}
\xi(E)=\frac{8\sigma_r}{\pi}\frac{E E_m}{\Delta^2}\frac{1}{(1-k\sigma_r)}\,\frac{1}{\Theta(1-k\sigma_r)}\,,\label{eq:sanchez2}
\end{equation}
where $\Theta(x)$ is the Heaviside function.
In comparison to the uncorrelated disorder there is then a faster increase of the localization length with energy and, at least at first order, a strict divergence at $k\sigma_r=1$. Careful studies that go beyond the Born approximation actually show that the localization is not completely destroyed above the effective mobility edge, but $\xi$ increases by at least on order of magnitude, depending on the disorder strength.

\begin{figure}[htp]
 \begin{center}
 \includegraphics[width=1\columnwidth]{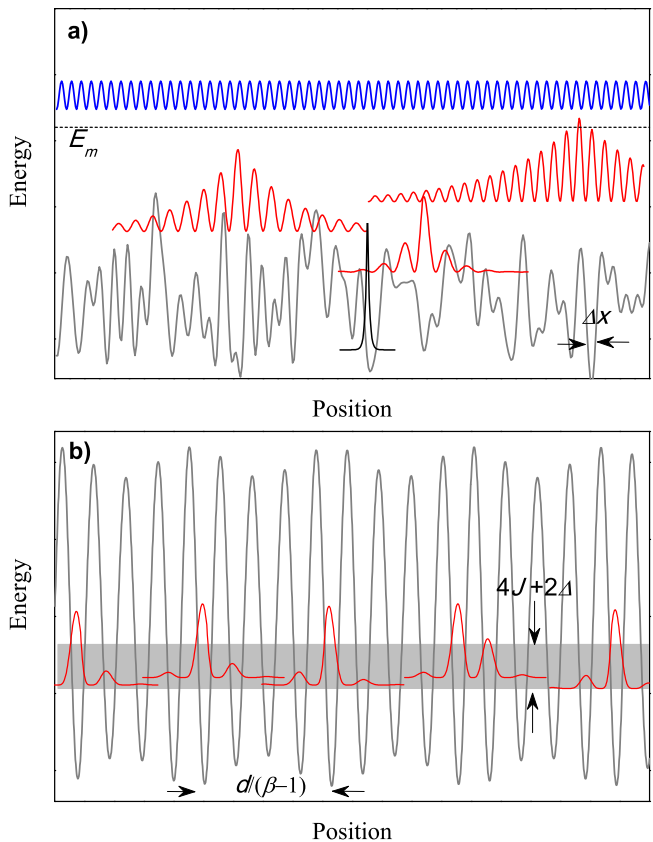}
\end{center}
 \caption{Cartoon of the spectrum of localized states in the two different systems that exhibit Anderson localization realized with Bose-Einstein condensates. a) Correlated random disorder in free space, created by laser speckles. The red states are exponentially localized, while the blue one is an extended state above the mobility edge and the black one is a strongly localized state in the Lifshits tail; b) Quasiperiodic lattice for $\Delta/J>2$. All the states related to the first energy band of the principal lattice (grey area) are exponentially localized. Here only the lowest set of states, separated on average by $d/(\beta-1)$, is shown.  }
 \label{fig:models}
\end{figure}

\subsection{Quasiperiodic lattices}
From the discussion above one could imagine that having an indefinite increase of the correlation length would lead to a vanishing of Anderson localization. However, there are special cases of potentials with non-decaying correlation functions that can support localized states, the price to pay being the need of a finite disorder strength for localization to occur. A paradigmatic example is the quasiperiodic lattice, that consists of a main lattice perturbed by a weaker lattice with a lattice constant that is incommensurate to the first one. In the absence of the perturbing lattice, the solutions of the single-particle problem are the well known extended Bloch states, that are formed by coherent superposition of the Wannier states of individual lattice sites. It is easy to imagine that if a non-periodic shuffling of the on-site energies is added, one might get into a situation where it is no longer possible to form extended Bloch states. As we outline in the following, theory shows indeed that above a critical strength of the secondary lattice the eigenstates of the system get exponentially localized as in random disorder. It is interesting to note that an analogous behavior appears in quasiperiodically driven systems, where localization occurs in presence of incommensurate driving frequencies, see for example ref.\cite{ringot00} and references therein.

Crystals with such quasiperiodic lattices, the so-called quasicrystals, are well known in solid-state physics \cite{quasicrystals}. The realization of quasiperiodic lattices for ultracold atoms is particularly simple \cite{Fallani07}. An individual sinusoidal lattice, the so-called optical lattice, can be created with the dipole potential of a laser beam arranged in standing wave configuration. A laser radiation with wavevector $k$ gives rise to a lattice with spacing $d=\pi/k_1$, and with a depth $U$ that is proportional to the laser intensity \cite{ovchi98}. The combination of a deep primary lattice of wavevector $k_1$ with a secondary one of incommensurate wavevector $k_2$ realizes the quasiperiodic lattice. A single particle in the first energy-band of a potential of this kind is actually a realization the well studied Harper \cite{harper55} or Aubry-Andr\'{e} model \cite{Aubry80}, which is presented in terms of the discrete Hamiltonian
\begin{eqnarray}
\nonumber H=&-J\sum_j (|w_j\rangle\langle w_{j+1}| + |w_{j+1}\rangle\langle w_j|)+\\
&+\Delta\sum_j\cos(2\pi\beta j)|w_j\rangle\langle w_{j}|\,.\label{eq:aa}
\end{eqnarray}
The first term describes the hopping of the particle to neighbouring sites of the primary lattice, with energy $J$, while the second contains the quasiperiodic shift of the on-site energies due to the secondary lattice. The kinetic term of the Hamiltonian has the very well known energy spectrum consisting in a whole set of extended Bloch states arranged in a band of energy width $4J$. The disordering term is parameterized by the wavevector ratio $\beta=k_2/k_1$ and the disorder amplitude $\Delta$, which is related to the potential depths of the two lattices by $\Delta\approx U_2/2\exp(-\beta^2/\sqrt{U_1/E_r})$, where $E_r=\hbar^2k_1^2/2m$ is the recoil energy \cite{Modugno09}. The incommensurability of the two lattices can be maximized by choosing $\beta$ as the ratio of two large consecutive elements of the Fibonacci series. This leads to the inverse golden ratio, $\beta=(\sqrt{5}-1)/2$. Note that in practice, in a finite-size system containing $N$ lattice sites it is unnecessary to have $\beta$ irrational. The weaker requirement for $\beta$ is instead that it can be expressed as a ratio of relatively prime natural numbers $\beta=P/Q$ such that $P,Q>N$.

It can be analytically demonstrated that the Aubry-Andr\'{e} model admits only extended states below a threshold disorder strength $\Delta=2J$, and only exponentially localized states above \cite{Aubry80,lifshits88}. As already discussed, the disordering potential in eq.\ref{eq:aa} has a non-decaying correlation function
\begin{equation}
g(z)=\Delta^2/2\pi\, \cos(2\pi\beta z)\,.\label{eq:aacorr}
\end{equation}
A localization length can however be defined also in this case, since the correlation is weak \cite{lifshits88}. A way to identify the value of the threshold $\Delta_t$ is to note that the Aubry-Andr\'{e} Hamiltonian is dual, i.e. one can find a transformation to the momentum space that gives rise to an Hamiltonian with an analogous form. The appropriate pseudo-momentum basis for the transformation is
\begin{equation}
|k_l\rangle=\sum_j\exp(i2\pi\beta k_l j)|w_j\rangle\,,
\end{equation}
and one readily gets the dual Hamiltonian
\begin{eqnarray}
\nonumber H=&-\frac{\Delta}{2J}J\sum_l (|k_l\rangle\langle k_{l+1}| + |k_{l+1}\rangle\langle k_l|)+\\
&+\frac{2J}{\Delta}\Delta\sum_l\cos(2\pi\beta l)|k_{l}\rangle\langle k_l|\,.\label{eq:aadual}
\end{eqnarray}
This second Hamiltonian has obviously localized states where the first one has extended states, and viceversa. Since the transition point from extended to localized states must be the same for the two, the only possibility is that this happens at $\Delta/2J=1$. Numerical solution of eq. \ref{eq:aa} for $\beta=(\sqrt{5}-1)/2$ shows indeed the presence of a very sharp transition between extended and localized states in an infinite system. Such transitions turns into a smoother crossover not only if the system has a finite extension, but also if $\beta$ is less irrational, as outlined above \cite{Modugno09}.

The localization length of the Aubry-Andr\'{e} model can be derived analytically from eq.\ref{eq:thouless} as $\xi=d/\ln(\Delta/2J)$, and is thus energy independent \cite{Aubry80}. Since localization is obtained only for $\Delta>2J$, except for a small region of disorder strengths $\Delta\approx 2J$, the localization length is comparable of smaller than $d$, hence quite short. The localization scenario for the first energy band of a quasiperiodic lattice above threshold, shown in Fig.\ref{fig:models} consists in $N$ equally-localized states that span an energy $2\Delta+4J$. Their energies keep memory of the quasi-periodicity and are therefore arranged in a sort of minibands structure \cite{Modugno09}. Note that there is not a Lifshits tail in quasiperiodic lattices, since all the localized states have a comparable overlap with the two neighbouring states.

It is important to stress again that the Aubry-Andr\'{e} model does not have a mobility edge, but just a critical disorder strength above which all states (in the first lattice band) are localized. It can be shown that a mobility edge appears instead if long-range tunnelling, e.g.tunnelling to the next-neighbouring sites, is allowed \cite{holtaus07,sarma09}. This is a regime that can be reached with ultracold atoms by employing relatively shallow optical lattices.

\subsection{Random disorder in a lattice}
It is now instructive to compare the two types of disordered systems above with the paradigmatic condensed-matter disorder model, i.e. the Anderson model \cite{Anderson58}. In the latter one considers a particle moving in a lattice with a random shift of the on-site energies. Simple physical systems that realize the Anderson model that have been investigated so far are microwave in guides \cite{izrailev,kuhl} and photonic lattices \cite{Schwartz07,lahini08}, but this is a configuration that can be explored also with Bose-Einstein condensates, for example employing a deep optical lattice that is perturbed by a weak speckle potential. Experiments with condensates are actually are underway \cite{White09}, although not yet in the regime of negligible interactions that we are discussing in this review.  The 1D Anderson model is represented by the discrete Hamiltonian
 \begin{equation}
H=-J\sum_j (|w_j\rangle\langle w_{j+1}| + |w_{j+1}\rangle\langle w_j|)+\sum_j\epsilon_j|w_j\rangle\langle w_j|\,,\label{eq:anderson}
\end{equation}
where the first term describes again the hopping and the second one contains the on-site energies $\epsilon_j$, which are supposed to be randomly distributed in the interval $[-\Delta,+\Delta]$. Assuming an uncorrelated random disorder, a perturbation theory approach gives in this case a localization length of the form \cite{kramer93}
\begin{equation}
\xi(E)\approx24d(4J^2-E^2)/\Delta^2\,,
\end{equation}
in the limit of $\Delta/J \ll 1$. The localization length is again inversely proportional to $\Delta^2$ as in free space, but the energy dependence is remarkably different, since the maximum of localization is obtained at the two band edges $E\approx \pm2J$, where the energy dependence is quadratic, $\xi(E)\propto E^2$. Comparison of this result with eq.\ref{eq:sanchez2} indicates that for low energies and the same speckle disorder, the localization length in a lattice is smaller than that in free space by a factor of the order of $E_m/4J$, because of the lower kinetic energy. It is interesting to note that considering the finite correlation length of the speckle potential might lead to the appearance of an effective mobility edge in 1D, in analogy with the free-space case \cite{izrailev}.  However, in the first band of the lattice the extended states appear only if the correlation length is larger than $d$, and they are located close to the band center, in analogy to what happens for the extended states above the true mobility edge of the 3D Anderson model.

\subsection{Anderson localization and weak nonlinearities}
Before discussing the experiments performed with disordered Bose-Einstein condensates, we need to briefly outline also the effect of a weak nonlinearity on Anderson localization. The fact that nonlinearities such as those arising from interactions between the particles can change the localization properties of a many-body system has been recognized since long, for example in condensed-matter systems \cite{kramer93}. The basic effect of a nonlinear interaction term in a disordered system is to provide a mean of mixing the localized single-particle states to give rise to new many-body states. In some cases this results only in a small modification of the localization properties of the system. There are notable cases where however a weak nonlinearity is sufficient to turn the eigenstates of the system from localized to fully extended, hence destroying Anderson localization. The details of how nonlinearities affect a disordered system depend of course from many details such as the quantum statistics of the particles (Bose or Fermi), the single-particle energy spectrum, the temperature of the system and the nature of the nonlinearity (attractive or repulsive, short- or long-ranged).

For atoms, the dominant interaction at ultralow temperatures is the $s$-wave contact interaction, which is parametrized by the scattering length $a$. The interaction energy per particle in a system of $N_{at}$ atoms in a state $\phi({\bf r})$ is \cite{dalfovo99}
\begin{equation}
E_{int}=\frac{4\pi\hbar^2}{m}\,aN_{at}\int|\phi({\bf r})|^4 d^3r\,.\label{eq:eint}
\end{equation}
Here a repulsion between atoms corresponds to a positive scattering length, i.e. $E_{int}>0$, while the opposite applies for attraction. In a standard Bose-Einstein condensate such interaction energy can easily be comparable or even much larger than the potential energy of the disorder. However, for some atomic species it is possible to precisely tune the scattering length from positive to negative values across a zero by employing Feshbach resonances \cite{chin09}. For example, in a magnetic Feshbach resonance the scattering length can be continuously modified simply by applying a homogeneous magnetic field that brings different internal states in resonance with the atomic state under consideration.

The case of most relevance for ultracold Bose-Einstein condensates is that of a repulsive interaction (i.e $a>$0), since for attractive interactions they are typically not stable. Let us then discuss intuitively the effect of an interaction energy that is weak, i.e. is comparable to the the potential energy of the disorder, but not much larger than the kinetic energy of the system. Let us assume that all the single-particle states of the disordered potential are localized, as it happens for example in quasiperiodic lattice above threshold. For zero interaction all the atoms will tend to condense in the localized ground state. A weak repulsion will instead force the bosons to occupy more than one of these states to minimize the system energy. These states can further mix between themselves to give rise to an extended collection of less-localized states if $E_{int}$ is comparable to their energy separation, or even to a fully extended system if $E_{int}$ is comparable to the full disorder amplitude $\Delta$. This intuitive picture will be further discussed in Section 4, in relation to experiments.
It is interesting to note that the effect of a weak interaction on a bosonic system is qualitatively different from that on a fermionic system. In the latter, a repulsion tends to keep the particles in their single-particle states, which are localized. An attraction on the contrary can favour the build-up of extended many-body states that span the whole sample, because this allows to establish an $E_{int}<0$ that lowers the system energy.

\section{Experiments on Anderson localization with Bose-Einstein condensates}
The idea of using Bose-Einstein condensates to explore Anderson localization appeared already several years ago \cite{Damski03,roth03}, but it took a while before all the required conditions for its observation in experiments could be met. Initially, the focus has been on random disorder in free space, created with laser speckles \cite{lye05,clement05,fort05,schulte05}. These pioneering experiments served to identify two different issues.

The first one was to realize disordered potentials that could give the proper range of localization lengths, i.e. smaller that the maximum system size, that is typically in the range of 1mm, while keeping the disorder strength to a minimum to avoid populating just strongly localized states in the Lifshits tail of the spectrum. For a speckle potential this condition requires a grain size $\sigma_r$ that is of the order of 1$\mu$m. Since this is the wavelength of the laser light used to create speckles, a proper design of the optical system is required. While some groups have worked in this direction \cite{clement06,Chen08}, others have chosen to work with quasiperiodic lattices, which naturally provide a very small $\xi$ \cite{lye07}.

The second issue was related to the unavoidable presence of interaction: all the pioneering studies have actually been performed with $^{87}$Rb atoms, which have a naturally large repulsion, parametrized by $a$=100$a_0$, and for which the magnetic control of the interaction is extremely difficult. The obvious way to solve this problem is to reduce the particles density, either by reducing the atom number \cite{lye07} or by letting the system expand to a large volume \cite{clement06}. However, the very weak interactions achievable in this way can have a non negligible  effect on the localization, and a real cancellation of the interaction can be achieved only via a Feshbach resonance, as we demonstrated for example with $^{39}$K atoms. Actually, while some initial experiments were indicating the onset of a localization regime, it has not been possible to achieve an unambiguous observation of Anderson localization of atoms in a Bose-Einstein condensate until two recent experiments in Paris \cite{Billy08} and Florence \cite{Roati08}.

\subsection{Localization in a speckle potential}
To observe Anderson localization the Paris team has combined a diffraction-limited lens system for the production of speckle potentials with small $\sigma_r$ \cite{clement06} with a reduction of the atomic density in a one-dimensional expansion of the sample in a waveguide. This has allowed a careful investigation of the diffusion of matter-waves in a random potential, with a detailed study of the localization length in 1D \cite{Billy08}.

The Paris experiment employs a Bose-Einstein condensate of $^{87}$Rb atoms, therefore naturally interacting. The condensate is initially prepared in a harmonic trap, and
subjected to speckle disorder with amplitude $\Delta/h=10-100$Hz and $\sigma_r$=0.26$\mu$m (this corresponds to a mobility edge at $E_m/h=h/2m\sigma_r^2\approx$ 850Hz). The radii of the sample in the trapping potential are about 35$\mu$m in the axial direction and 3$\mu$m in the two transverse directions. The corresponding $E_{int}$ is as large as $h\times$220 Hz, hence definitely non negligible. The idea was then to let the condensate suddenly expand out of the trap into a one-dimensional optical waveguide aligned along the weak axis of the trap and transversely to the direction of propagation of the speckles beam. During the initial stage of the expansion $E_{int}$ is converted into kinetic energy, the condensate rapidly diffuses out of the initial tight spatial distribution, and gradually becomes almost non-interacting as its density drops with the expansion. At a later stage its wavefunction gets to be a superposition of the single-particle localized eigenstates of the disordered potential, and the motion therefore is confined in space. In particular, the decay of the tails of the atomic distribution is determined by the localization properties of the largest momentum components, which have an energy $E\approx E_{int}$.  To get into the regime where trivial localization in deeply bound states cannot be possible, care has been taken to have $E_{int}>\Delta$, while the maximum energy has been kept below the effective mobility edge of eq.\ref{eq:sanchez2} by having $E_{int}<E_m$.

\begin{figure}[htp]
 \begin{center}
 \includegraphics[width=1\columnwidth]{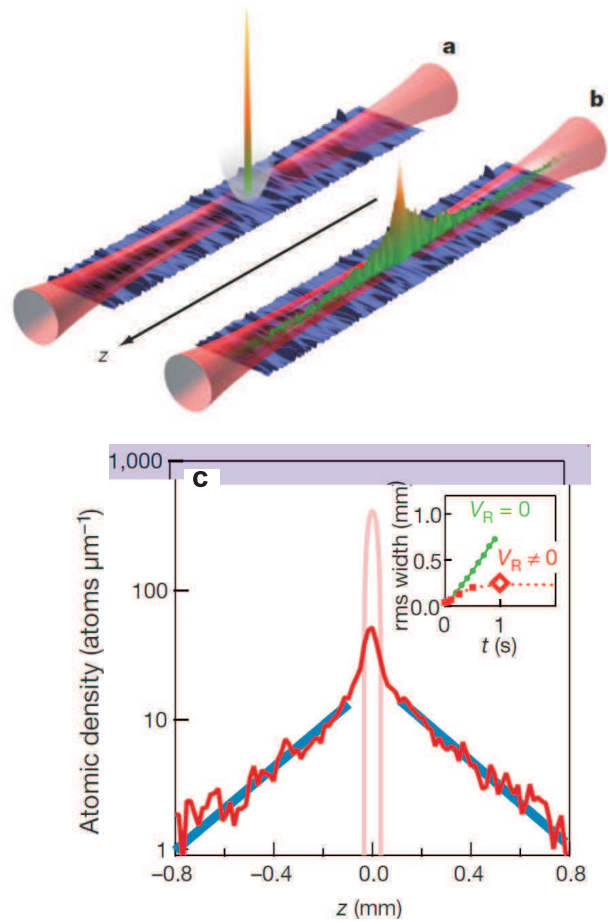}
\end{center}
 \caption{Geometry of the localization experiment in random disorder: a) an interacting Bose-Einstein condensate is initially prepared in a harmonic trap; b) the condensate is then let free to expand into a 1D guide, where it is exposed to the speckle potential. c) Experimental observation of localization: for sufficiently large disorder strengths the condensate expansion stops, and the tails of its density distribution acquire an exponential decay behaviour. Figure reprinted with permission from \cite{Billy08}. Copyright 2010 by Macmillan Publisher Ltd.}
 \label{fig:paris1}
\end{figure}

Fig. \ref{fig:paris1} reports the typical experimental observation. The axial density distribution of the atomic sample after some expansion in the guide is measured by absorption imaging. For vanishing disorder strengths $\Delta$ the localization length of eq.\ref{eq:sanchez2} is much larger than the size of the observation region $L$ and the expansion appears to be ballistic, i.e the rms size of the sample increases linearly with time. As $\Delta$ is increased above a critical strength such that $\xi(E_{int})<L$ the size of the sample stops to increase after some initial expansion, and the atomic density distribution acquires a clear exponential decay character in its tails. Note that a fraction of the atoms remains trapped in the central region with the typical inverted-parabola shape of the original condensate. This is presumably a consequence of a trivial localization of a fraction of the sample in more deeply bound states in the Lifshits tail of the spectrum, $E<\Delta$, something that was observed already in previous experiments \cite{lye05,clement05,fort05,schulte05}.

\begin{figure}[htp]
 \begin{center}
 \includegraphics[width=0.9\columnwidth]{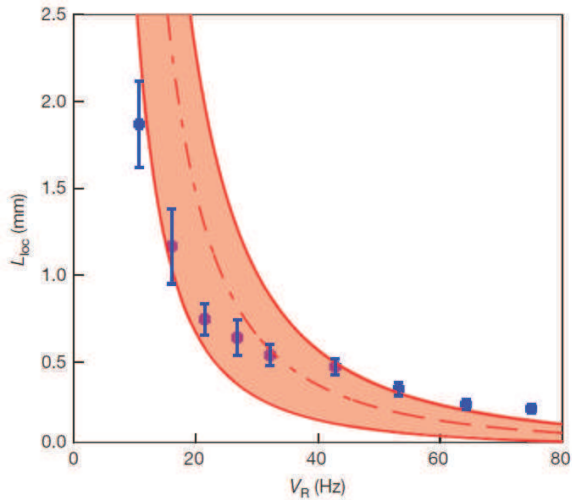}
\end{center}
 \caption{Dependence of the measured localization length on the disorder strength. The experimental data (blue dots) are in good agreement with the prediction of theory (dashed-dotted line). The red region accounts for the uncertainty in the determination of $E_m$ and $E_{int}$ in the experiment. Figure reprinted with permission from \cite{Billy08}. Copyright 2010 by Macmillan Publisher Ltd.}
 \label{fig:paris2}
\end{figure}

The Paris group has then studied the $\Delta$-dependence of the localization length, which is extracted from an exponential fit of the tails of the distribution to a function of the kind $n(x)=n_0\exp-|x|/\xi$. The experimental observations shown in Fig.\ref{fig:paris2} feature the expected decrease of $\xi$ for increasing $\Delta$ and very positively compare to the prediction of eq.\ref{eq:sanchez2}, integrated over the density of states \cite{sanchez07}.

The regime of energies that are partially above the mobility edge $E_m$ has also been studied in the experiment, by increasing the initial interaction energy to have $k\sigma_r>1$. In this case the largest momentum components are not localized and expand indefinitely, producing a depletion of the extreme tails of the density distribution. The experimental observation is indeed of a faster decay of the remaining atomic distribution, which as an algebraic dependence $n(x)=n_0/|x|^\alpha$, with $\alpha\approx$2, as predicted by theory \cite{sanchez07}.

A natural question is now whether the reduction of the interaction energy that is achieved through the expansion is sufficient to neglect completely its effect on the localization properties. If the measured localization length in Fig.\ref{fig:paris2} is used to estimate the particle density, then one finds a reduction of the interaction energy by approximately a factor of 100 for the largest disorder, or 3000 for the smallest one. This brings the initial interaction energy to about 1Hz or below, hence almost two orders of magnitude lower than the mean disorder energy. While such a small interaction energy will probably not affect the localization properties on the relatively short time scale of the present experiment, as we will discuss later its effects might be still visible in the long-time dynamics of the system.

\subsection{Localization in a quasiperiodic lattice}
In Florence we have instead employed a Bose-Einstein condensate of $^{39}$K atoms, where the interaction can be almost nulled via a broad zero-crossing close to a Feshbach resonance \cite{Roati07,derrico07,fattori08}, in combination with a 1D quasiperiodic lattice. With this approach the observation of the localization phenomena does not rely on a dynamical expansion of the sample, allowing the investigation of both dynamical and static properties of the localized condensate. The study of the spatial and momentum distributions has allowed to verify the expected localization properties of a quasiperiodic system described by the Aubry-Andr\'{e} model.

\begin{figure}[htp]
 \begin{center}
 \includegraphics[width=1\columnwidth]{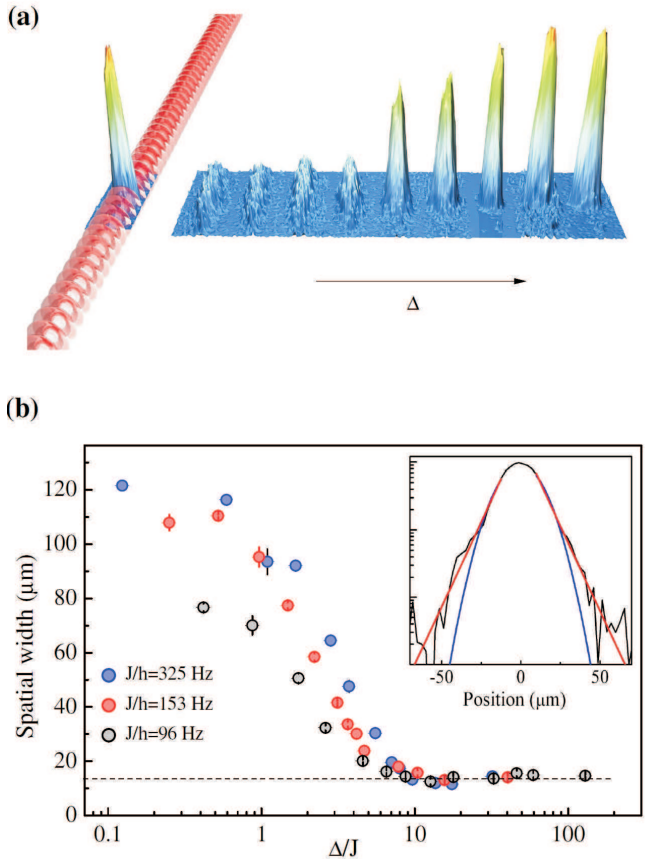}
\end{center}
 \caption{Diffusion of a non-interacting Bose-Einstein condensate in a quasiperiodic lattice. a) Cartoon of the experimental realization. An initially trapped condensate (left) is let free to expand in an optical guide along which is arranged the quasiperiodic lattice. The size of the condensate after a fixed expansion time (right) depends on the strength $\Delta$ of the secondary lattice, and progressive localization is observed for increasing $\Delta$. b) Axial size of the condensate after 700 ms of expansion, for three different values of the tunneling energy $J/h$. The inset shows the exponential decay of the tails in the localized regime: the blue curve is a gaussian fit, the red curve an exponential fit. Figure reprinted with permission from \cite{Roati08}. Copyright 2008 by Macmillan Publisher Ltd.}
 \label{fig:diffusion}
\end{figure}

In the experiment an interacting BEC is initially prepared in a harmonic potential \cite{Roati08}. A tight 1D lattice of wavevector $k_1$ is then raised using a slow double-exponential ramp, optimized to keep the system in its ground state; the lattice beams provide also a radial confinement, thus realizing a guide for the atoms. A second weaker lattice with incommensurate wavevector $k_2$ is then raised with a similar ramp lasting about 100 ms, i.e. a time longer than $h/J$. The scattering length is eventually reduced to a very small value $a\approx0.1a_0$, which corresponds to a residual interaction energy that is negligible in comparison with the kinetic energy ($E_{int}<0.01J$). The ratio of the two wavevectors is $\beta=1032\rm{nm}/862\rm{nm}\approx1.1972$. In a first stage, this set-up has allowed to study the spatial diffusion of the condensate along the guide, similarly to the Paris experiment. A main difference is however that in this case the diffusion is driven only by the kinetic energy associated to the trapping potential. The typical diffusion dynamics is shown in Fig.\ref{fig:diffusion}: a) when $\Delta=0$ the motion is ballistic; b)when $\Delta<2J$ the motion is still ballistic, but with a reduced speed, (because of the reduced tunnelling energy in the minibands of the superlattice); c) when $\Delta>2J$ no motion is discernible, in accordance with the expectation of localized states with $\xi<d$, which is in turn much smaller than the initial size of the system, $L\approx20 d$. Interestingly, the possibility of tuning  $J$ independently from $\Delta$, allowed also to confirm that such change of diffusion dynamics has the proper scaling with $\Delta/J$, as expected from eq.\ref{eq:aa}.
The analysis of the tails of the spatial distribution along the quasiperiodic lattice has allowed instead to detect the appearance of an exponential localization of the far tails of the density distribution for $\Delta>2J$.

In a second stage we have studied the momentum distribution $\rho(k)$ of the stationary state of the system in the harmonic trap. This is obtained by suddenly releasing the sample from all potentials, allowing a long ballistic expansion driven by the initial momentum, and eventually taking an image of the final spatial distribution along the lattice direction. An inspection of the evolution of the momentum distribution for increasing $\Delta$ elucidates the peculiar localization mechanism of quasiperiodic lattices. In absence of disorder one has the usually narrow $\rho(k)$ corresponding to the $q\approx 0$ quasimomentum states of an ordered lattice. For weak disorder, with $\Delta<2J$, two momentum components appear in the first Brillouin zone, at $2k_2-2k_1=2(\beta-1)k_1$ and $4k_1-2k_2=2(2-\beta) k_1$, which are the two Fourier components of the $\cos(2\pi\beta j)$ term in eq.\ref{eq:aa}. As $\Delta$ increases, other smaller components arise, until the whole first Brillouin zone is occupied for $\Delta\geq2J$, implying the presence of states localized over a distance $\xi<d$, as expected.

\begin{figure}[htp]
 \begin{center}
 \includegraphics[width=1\columnwidth]{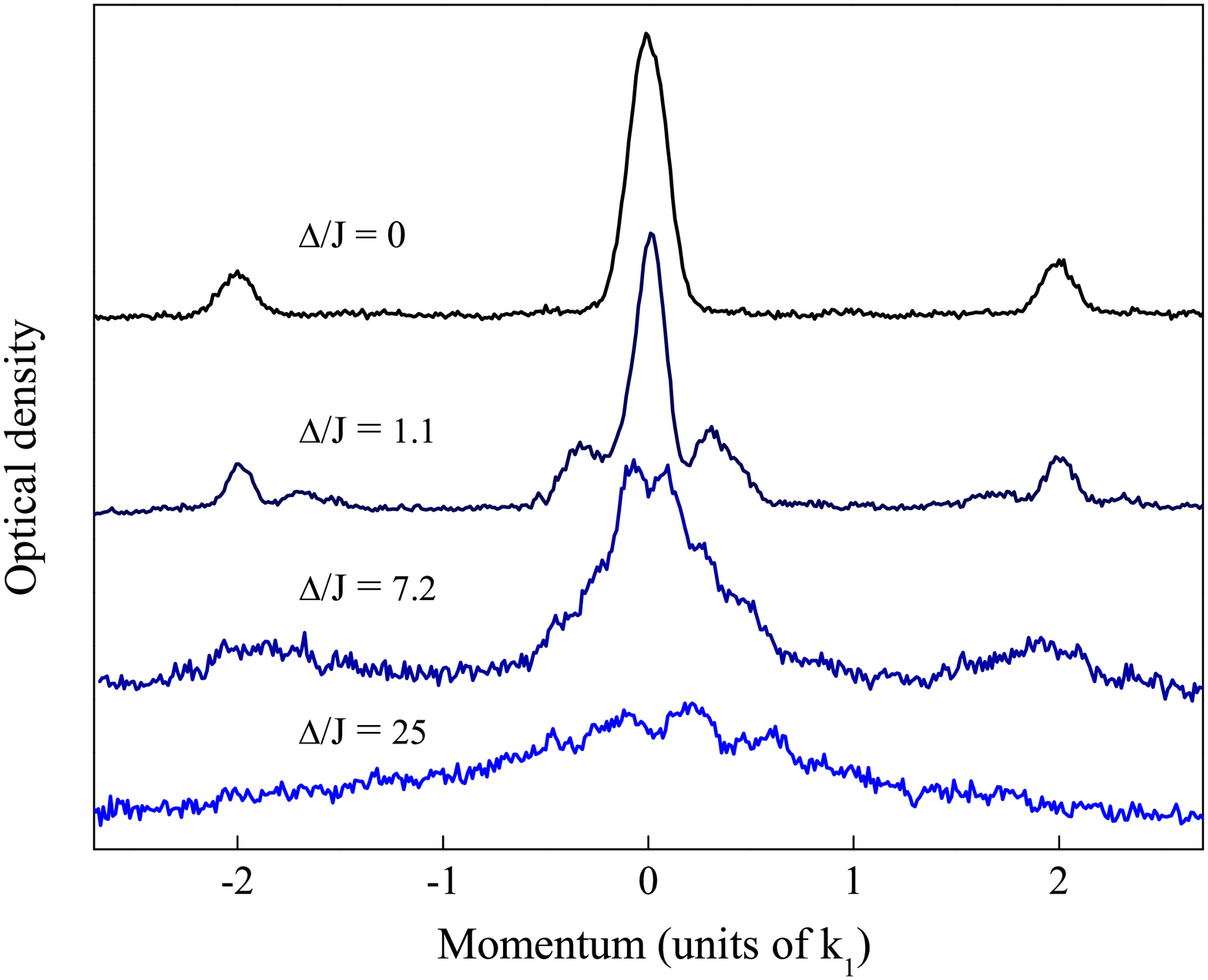}
\end{center}
 \caption{Stationary momentum distribution of a noninteracting Bose-Einstein condensate in a quasiperiodic lattice, for increasing strengths of the disordering lattice. The broadening of the distribution provides evidence of a progressive localization.}
 \label{fig:momentum}
\end{figure}

The momentum distribution gives also information about the quasiperiodic spatial arrangement of the localized eigenstates of the system. When reducing the size of the system in the localized regime of $\Delta>2J$ by increasing the harmonic confinement, an interference with period $\delta k=2(\beta-1)k_1$ clearly appears. This is due to the interference of localized states belonging to the lowest-lying energy miniband \cite{Modugno09}, that are indeed separated on average by $d/(\beta-1)\approx5.1 d$. The phase of such interference varies randomly during multiple repetitions of the experiment, indicating that the localized states have indeed not a fixed phase relation, i.e. they are independent. Actually, if the number of states is increased above 3-4 the interference contrast rapidly vanishes, as expected if the relative phases are random. The occupation of several localized states testifies that the system is not in its ground state, despite the preparation procedure has been designed to minimize the excitations. This is probably due to the diverging tunnelling time between states at the onset of the localization, a common issue in the preparation of quantum systems that feature insulating phases \cite{edwards08,zakrewski09}.

\begin{figure}[htp]
 \begin{center}
 \includegraphics[width=1\columnwidth]{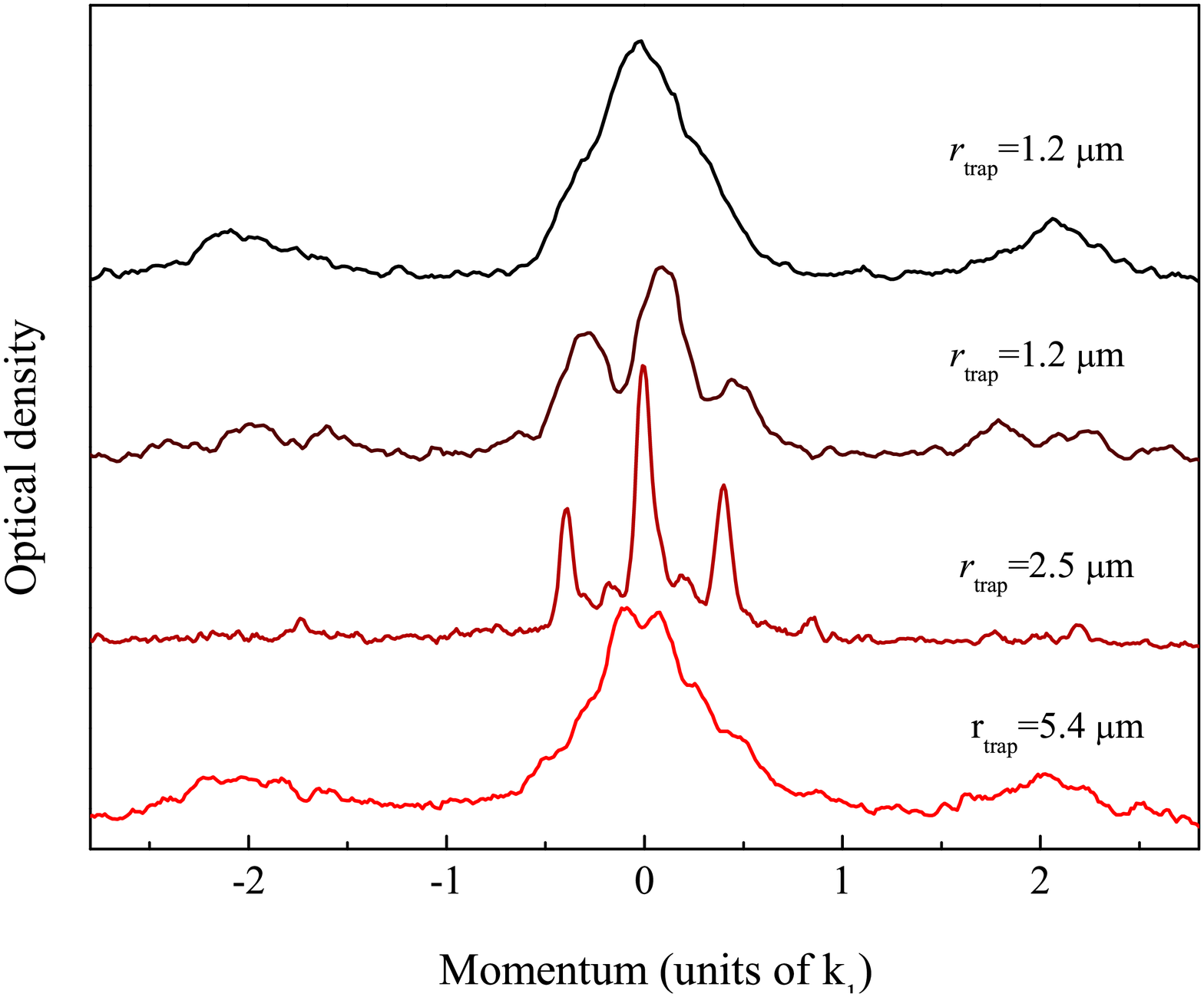}
\end{center}
 \caption{Counting the localized states of a Bose-Einstein condensate in a quasiperiodic lattice. Momentum distribution for varying sizes of the harmonic trap ($r_{trap}=\sqrt{\hbar/m\omega}$), corresponding to about: a) one state; b) two states; c) three states; d) seven states. In the latter case the interference vanishes, since the states are phase incoherent.}
 \label{fig:glass}
\end{figure}

It is interesting to note that, while the the Aubry-Andr\'{e} model has been confined to theory for several decades, soon after the experimental realization with Bose-Einstein condensates an analogous experiment has been performed at the Weizmann Institute with photons propagating in a quasiperiodic \emph{photonic} lattice \cite{lahini09}. Here a narrow light wavepacket is injected into one side of the quasiperiodic lattice, and the evolution of its transverse profile is monitored along the propagation direction. The localization transition has been detected by studying a quantity that if often employed to quantify the spread of a disordered system, the so-called participation ratio (PR): PR= $(\sum_n|\phi_n|^2)^2/\sum_n|\phi_n|^4$. A sharp variation of the participation ratio has been indeed observed when varying the strength of the disordering modulation of the photonic lattice, centered at experimental parameters corresponding to $\Delta/J=2$ in eq. \ref{eq:aa}.
\begin{figure}[htp]
 \begin{center}
 \includegraphics[width=0.8\columnwidth]{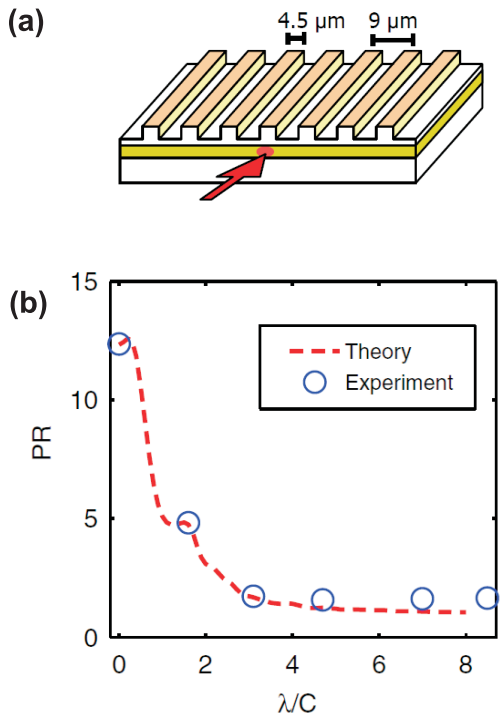}
\end{center}
 \caption{Transverse localization on light in a quasiperiodic photonic lattice. a) Geometry of the experiment. b) Evolution of the participation ratio with the strength of the lattice modulation. Here the parameter $\lambda/C$ is equivalent to the parameter $\Delta/J$ of the atomic case. Figure reprinted with permission from \cite{lahini09}. Copyright 2009
by the American Physical Society.}
 \label{fig:photonic1}
\end{figure}

\section{Experiments on the interplay of disorder and nonlinearities}
As already mentioned, ultracold quantum gases with magnetically tunable interaction offer also the possibility of studying the effect of nonlinearities on Anderson localization. In the following we briefly review a recent experiment with a BEC in quasiperiodic lattices in this direction, and we relate the findings to those of experiments with photonic systems.

Let us start by repeating in detail the reasoning of Section II.D. We can therefore imagine to have an ideal Bose-Einstein condensate that is prepared at $T=0$ in the ground-state of a quasiperiodic lattice with $\Delta$ above the localization threshold. The system is therefore in the regime of Anderson localization. The addition of a weak repulsion can instead drive the system into at least two different regimes.
If the repulsion is so weak that the average interaction energy per particle $E_{int}$ is smaller than the characteristic disorder energy $\Delta$, its main effect is just to prevent the occupation of the ground state alone. The system energy is indeed minimized by distributing the particles in more than one single-particle state, so that the interaction energy arising from eq.\ref{eq:eint} is reduced. In this case, one can imagine that the various states populated by the condensate keep their single-particle character and are therefore still localized. This regime is often indicated in condensed-matter theory as that of an Anderson glass \cite{Giamarchi88}, since it features the main properties expected for a glass. First of all, this is an insulating regime, since the system is composed of spatially localized states. Second, the energy cost for creating excitations, i.e. for moving particles from one localized state to another, vanishes as the system size tends to infinity, since the energy to be compensated is just the one due to disorder.

For increasing repulsion one can expect that the number of localized states that are populated increases. When $E_{int}\sim\Delta$ there will be a high chance that states that are neighbouring in space become also neighbouring in energy on the scale of $E_{int}$. In this case groups of neighbouring localized states start to mix and give rise to less localized states, or eventually to fully extended states. One might also describe this delocalization as the results of a screening of disorder by the interaction \cite{lee92}. One can indeed imagine that most of the particles are arranged in the single-particles states to provide a spatially-varying interaction energy that almost completely compensates the disordered potential, while the few remaining particles can freely move on top of such combined potential. For identical bosons this picture corresponds to a fully delocalized system. A not too weak repulsive interaction therefore is expected to keep bosons delocalized, while the opposite, i.e. a stronger localization, would happen for an attractive interaction.

\begin{figure}[htp]
 \begin{center}
 \includegraphics[width=1\columnwidth]{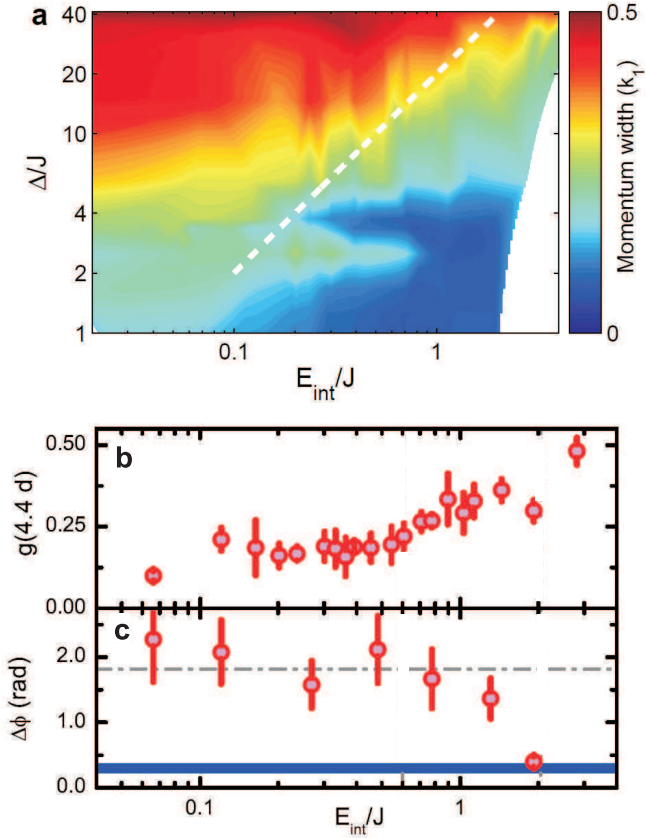}
\end{center}
 \caption{Delocalizing effect of a repulsive interaction on a Bose-Einstein condensate in a quasiperiodic lattice. a) Root-mean-squared width of the momentum distribution for varying $\Delta$ and $E_{int}$: the interaction tends to make the system more extended (narrower momentum width) or, alternatively, to increase the threshold $\Delta/J$ for localization. The white line is the value of $E_{int}$ that is necessary for delocalization as derived from a screening argument. The correlation between neighbouring localized states (b) increases for increasing interaction, while their phase fluctuation (c) decreases, and finally reaches the one measured below the localization threshold (blue area). Figure reprinted with permission from \cite{deissler09}. Copyright 2010 by Macmillan Publisher Ltd.}
 \label{fig:aainteract}
\end{figure}

To verify this conjecture, we have studied how the $\rho(k)$ changes when adding a repulsive interaction, by means of the same Feshbach resonance that used to cancel the natural interaction. The experimental techniques are the same of Section III.B, except for the fact that during the loading in the lattice the scattering length is no more reduced to zero, but kept to some positive value in the range $a=1-400 a_0$. Fig. \ref{fig:aainteract}, shows the evolution of the width $\rho(k)$ in the $\Delta-E_{int}$ plane.

The trend of the data in Fig.\ref{fig:aainteract} a is clear: a system that is localized for vanishing $E_{int}$ ($\rho(k)$ is broad) tends to become more extended as the interaction energy is increased at fixed $\Delta$. Eventually, for not too large $\Delta$, the momentum distribution goes back to that of a condensate below the localization thershold. Alternatively, the diagram can be interpreted as a shift to larger $\Delta$ of the localization threshold for increasing repulsion. Other information can be extracted from the profiles of $\rho(k)$. For example, all the states that are initially localized belong to the first miniband of the spectrum, which has a width $\Delta'\ll\Delta$. As soon as $E_{int}$ becomes equal to $\Delta'$ (white line in Fig.\ref{fig:aainteract} a), a restoring of the interference modulation that signals the reestablishing of the phase coherence between neighbouring states is observed. To quantify the correlation between the states, one can evaluate the spatially-averaged correlation function of the condensate $g(z)=\int dx \psi^{\ast}(x)\psi(x+z)$, which is directly related to the Fourier Transform of $\rho(k)$ by the Wiener-Kintchine theorem: $g(z)=F^{-1}\rho(k)$. The natural distance for monitoring the evolution of the correlations in a quasiperiodic lattice is of course the mean separation between the lowest laying states, $z=d/(\beta-1)$. Alternatively, one can measure the fluctuations of the phase of the interference pattern in the momentum distribution, by repeating the experiment many times under the same conditions \cite{deissler09}. Both these quantities are shown in Fig.\ref{fig:aainteract} b-c for one particular value of $\Delta/J\approx 12$ above the localization threshold. In the Anderson localization regime $g(z)$ is almost zero, while the phase fluctuations are consistent with having totally random phases in the system. As $E_{int}$ increases there is a first regime of $E_{int}<\Delta'$ in which the two observables do not change appreciably, confirming that the only effect of the repulsion is to promote the occupation of a larger number of localized states. Only when $E_{int}\geq\Delta'$ the correlations increase and the phase fluctuations decrease, eventually reaching the background value measured for a Bose-Einstein condensate below the localization threshold.

These observations are in nice quantitative agreement with a theoretical study of the ground state of the quasiperiodic lattice \cite{deissler09}, and also in qualitative agreement with the general phase diagrams derived for random disorder \cite{Lugan07,natterman09,savona09} and quasiperiodic lattices \cite{minguzzi09}.

One important question at this point is about the dynamical properties of the extended system that is recovered for large repulsions and $\Delta>2J$: would the system expand indefinitely if the confinement is removed, as it would do in the superfluid regime $\Delta<2J$? This question has been long debated in theory \cite{shepelianski93,pikovsky08,flach09,larcher09}, where a variety of disorder models have been explored mainly by numerical methods. Recent studies \cite{pikovsky08,flach09,larcher09} agree in indicating that an interacting, disordered bosonic system in 1D should exhibit an indefinite expansion if the interaction energy is initially large enough to mix neighbouring localized states. The expansion however happens in a sub-diffusive manner, with the size of the system increasing with time approximately as $L(t)=A(t_0+t)^{\alpha}$, where the exponent $\alpha$ is not only smaller than unity, as one would expect for a ballistic expansion, but it is even smaller than the value 0.5 expected for normal diffusion.. This is a consequence of the fact that the states that are instantaneously mixed are limited in number and their number decreases with time as the interaction energy decreases with the progressive expansion, see for example ref.\cite{flach09}. The expectation is of a rather small exponent, $\alpha\approx0.2$ that should be rather independent from the interaction energy and the disorder strength. The subdiffusive character of the dynamics implies a very long time scale for observing a strong variation of the size of the system, that for our quasiperiodic lattice amounts to several seconds \cite{larcher09}. However, experiments are on the way to test the predicted subdiffusive expansion behavior. One important aspect predicted by theory is that the subdiffusion should continue indefinitely if the initial interaction energy is large enough to provide delocalization; this clearly rises once more a question about the very long time dynamics that could be observed also in the Paris experiment.

The possibility of studying in a controlled way the interplay of disorder and a widely tunable interaction is for sure the most interesting novelty introduced by ultracold quantum gases on the scenario of disordered systems. A similar capability was previously offered only by photonic systems. One example is the propagation of light through nonlinear atomic media \cite{chaneliere04,wellens05,shatokin05}. Another very interesting example is the transverse localization of light in  photonic lattices, where nonlinearities can be introduced thanks to the Kerr effect and controlled independently from the other system's parameter by choosing both an appropriate value of the nonlinear index of refraction of the material and the light intensity \cite{Schwartz07,lahini08,lahini09}. The kind of nonlinearity that has been explored so far in experiments is self-focusing, which corresponds to an attractive interaction in the atomic case. Also defocusing nonlinearities can be in principle explored \cite{Schwartz07}, although the strength of the nonlinear term in photonic systems is much smaller than the one for atoms, and the study of focusing nonlinearities is highly favoured. In this sense photonic lattices allow investigations that are complementary to those with Bose-Einstein condensates. Another complementarity of the two kinds of systems stays in the initialization of the system: while in quantum gases the initial state is typically a broad wavepacket that has overlap with many localized eigenstates, the macroscopic size of photonic lattices allows also to take as initial state almost exactly a single localized mode, or even a single lattice site.

The effect of nonlinearities on the transverse localization of light has been tested so far in both randomly disordered 1D and 2D lattices that realize the Anderson model \cite{Schwartz07,lahini08} and quasiperiodic lattices \cite{lahini09}. All the observations can be interpreted in the frame of the heuristic argument of Section 2.D: in the regime where the linear waves would be localized, the self-focusing nonlinearity mixes localized modes that are close in energy and space, providing in general a delocalization of the otherwise localized modes \cite{lahini08,lahini09}. If the linear waves are instead not yet localized, either because the $\Delta/J$ parameter of the of the quasiperiodic lattice is below threshold or the localization length in a random lattice is larger than the system size, then the effect of the nonlinearity is to promote localization. This is in qualitative agreement with the observed increase of the threshold for a BEC with repulsive interaction \cite{deissler09}. A very interesting study in a random 1D lattice has shown that the effect of the "attractive" nonlinearity is opposite on the lowest localized modes of the band with respect to the highest modes. In the former case it leads to an increase of the localization \cite{lahini08}, something that can be understood in terms of a negative $E_{int}$ that promotes the occupation of the absolute lowest localized modes and shifts them out of the band. The partial delocalization that is instead observed in the latter case can be justified not only as a shift of the initially localized states towards the center of the band, where the states are less localized, but also as the result of an effective "repulsive" interaction that results from the particular phase structure of the lattice modes in the top of the band.

\begin{figure}[htp]
 \begin{center}
 \includegraphics[width=0.95\columnwidth]{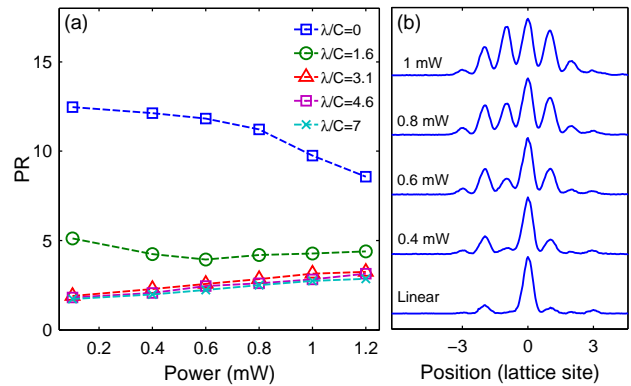}
\end{center}
 \caption{Effect of a self-focusing nonlinearity on the transverse localization in a quasiperiodic photonic lattice. a) Evolution of the participation ratio (PR) with the strength of nonlinearity, which is determined by the power level of the propagating light itself. Below the localization threshold ($\lambda/C$=2) the nonlinearity promotes localization, while above it promotes delocalization. b) Delocalization of the wavepacket for increasing nonlinearity at $\lambda/C$=3. Reprinted with permission from \cite{lahini09}. Copyright 2009 by the American Physical Society. }
 \label{fig:photonic}
\end{figure}

\section{Outlook}
We have so far discussed how Bose-Einstein condensates can be employed to provide a direct measurement of the localization of weakly interacting systems in one-dimensional random and quasiperiodic potentials. The two kinds of systems provide complementary information on the mechanism of Anderson localization and on its interplay with interactions. The possibility of tuning the atom-atom interaction in a controlled way through Feshbach resonances is an invaluable tool to evidence Anderson localization and to explore in a quantitative way the effect of nonlinearities. Obviously, also ultracold Fermi gases can be employed for this kind of studies; also there the interaction can be magnetically controlled, or can even be switched off by using a sample of identical fermions. Let us now outline some of the disorder-related phenomena that have been debated over the last few decades, and whose details could be finally disclosed in upcoming experiments with quantum gases, in both regimes of weak and strong interactions.

On the \emph{weak} interaction side, both Bose and Fermi gases might allow to extend the study of Anderson localization in random potentials to higher dimensionality, by means of a rather straightforward extension of the techniques employed so far. For example, there is particular interest in studying the characteristics of the mobility edges at the crossover between 2D and 3D in presence of a controlled interaction \cite{kuhn,shapiro}. This kind of studies would be interesting also in relation to the observed metal-insulator transition in disordered 2D electron gases \cite{lee}. First experiments in 2D disordered systems have actually been started \cite{bourdel}, and others will soon follow also in 3D. A question of general interest is also how much superfluidity or superconductivity persists in presence of disorder. Theoretical studies in this direction with Bose-Einstein condensates \cite{pavloff} and with Fermi gases in the BEC-BCS crossover \cite{orso,pezze09,demelo} are on the way, and future experiments might provide insight in the physics of "dirty" Bose and Fermi liquids that is relevant to various condensed-matter systems.

A particularly important question that is directly related to the ongoing experiments is about the nature of the many-body excitations of an interacting disordered system. A theoretical finding is that a system with an extended ground state might have excitations which are localized by disorder, although with a $\xi$ different from the the single-particle ones \cite{sanchez06}. Recent studies show also that such "many-body localization" may be destroyed is the density of excitations are increased above a threshold by increasing the temperature of the system. This should lead to a temperature-driven metal-insulator transition in both Bose \cite{aleiner09} and Fermi \cite{altshuler06} systems.

The other interesting limit is the one of interactions that are \emph{strong} enough to drive a lattice system into the regime of strong correlations, a regime that we did not discuss at all in this review. Here already in absence of disorder the interaction alone is capable of bringing the system into the Mott insulating phase, as observed in both Bose \cite{mottb} and Fermi gases \cite{mottf,mottff}. Despite extensive theoretical studies, there are still open questions about the interplay of disorder and interaction in this regime, even for the bosonic case. For example, at the Mott transition the disorder can in principle have two opposite effects, either a destabilization of the Mott insulator towards the superfluid, or a destabilization of the superfluid towards a new insulating phase, the so-called Bose glass. In theory there's still debate about the exact shape of the phase diagram \cite{Giamarchi88,Fisher88,Fisher89,roscilde08,roux08}. The experimental investigation of the phase diagram of these systems has just started \cite{Fallani07,White09,demarco09} (see \cite{Fallani08} for a review). Analogous phenomena are of course expected in the fermionic case. With the refinement of experimental and theoretical techniques, the study of even more fragile disorder-related phenomena appears to be within reach with ultracold quantum gases, such as quantum spin glasses and disordered magnetic phases \cite{palencia09}.

The author acknowledges enlightning discussions with Yoav Lahini and with the colleagues in the Quantum gases group at LENS, in particular with Benjamin Deissler, Massimo Inguscio, Michele Modugno, Giacomo Roati and Matteo Zaccanti.

\end{document}